\documentclass[letter]{aa}

\usepackage{txfonts}
\usepackage{graphicx}

\begin{document}

   \title{Strong radial segregation between sub-populations of evolutionary homogeneous stars in the Galactic
globular cluster NGC 6752.
   \thanks{Based on observations with the 1.3 m Warsaw telescope at Las Campanas
Observatory}}

      \author{V. Kravtsov\inst{1,2}
          \and
              G. Alca\'ino\inst{3}
          \and
              G. Marconi\inst{4}
           \and
              F. Alvarado\inst{3}
              }

\offprints{V. Kravtsov}

   \institute{Instituto de Astronom\'ia, Universidad Cat\'olica del Norte,
              Avenida Angamos 0610, Antofagasta, Chile\\
              \email{vkravtsov@ucn.cl}
            \and
              Sternberg Astronomical Institute, University Avenue 13,
              119992 Moscow, Russia\\
            \and
              Isaac Newton Institute of Chile, Ministerio de Educaci\'on de Chile,
              Casilla 8-9, Correo 9, Santiago, Chile\\
              \email{inewton@terra.cl, falvarad@eso.org}
            \and
              ESO - European Southern Observatory, Alonso de Cordova 3107, Vitacura,
              Santiago, Chile\\
              \email{gmarconi@eso.org}
             }

   \date{Received xxxxx / Accepted xxxxx}

   \abstract
{} {We investigate the new and still poorly studied matter of
so-called multiple stellar populations (MSPs) in Galactic globular
clusters (GGCs). Studying MSPs and their accumulated data can shed
more light on the formation and evolution of GGCs and other closely
related fundamental problems. We focus on the strong relation
between the radial distribution of evolutionary homogeneous stars
and their $U$-based photometric characteristics in the nearby GGC
NGC 6752 and compare this with a similar relation we found in NGC
3201 and NGC 1261.} {We use our new multi-color photometry in a
fairly wide field of NGC 6752, with particular emphasis on the $U$
band and our recent and already published photometry made in NGC
3201 and NGC 1261.} {We found and report here for the first time a
strong difference in the radial distribution between the
sub-populations of red giant branch (RGB) stars that are bluer and
redder in color $(U-B)$, as well as between sub-giant branch (SGB)
stars brighter and fainter in the $U$-magnitude in NGC 6752.
Moreover, the fainter SGB and redder RGB stars are similarly much
more centrally concentrated than their respective brighter and bluer
counterparts. Virtually the same applies to NGC 3201. We find
evidence in NGC 6752 as in NGC 3201 that a dramatic change in the
proportion of the two sub-populations of SGB and RGB stars occurs at
a radial distance close to the half-mass radius, $R_h$, of the
cluster. These results are the first detections of the radial trend
of the particular photometric properties of stellar populations in
GGCs. They imply a radial dependence of the main characteristics of
the stellar populations in these GGCs, primarily of the abundance,
and (indirectly) presumably of the kinematics.} {}

   \keywords {globular clusters: general --
                globular clusters: individual: MGC 6752, NGC 3201, NGC 1261}

\maketitle

\section{Introduction}
\label{introduc}

The southern Galactic globular cluster (GGC) NGC 6752 is one of the
nearest and most frequently studied GGCs. However, we here deal with
a fairly new aspect, which is related to photometric manifestations
of the so-called multiple stellar populations (MSPs) in GGCs. Beside
$\omega$ Cen they were initially revealed thanks to accurate Hubble
space telescope photometry in most massive GGCs, such as NGC 2808
(Piotto et al. \cite{piottoetal07}) NGC 1851 (Milone et al.
\cite{milonetal08}) among others, caused by the splitting of the
sub-giant branch (SGB) or/and of the main sequence (MS). The
splitting of the red giant branch (RGB) in most massive GGCs was
also shown by Lee et al. (\cite{leetal09}) using ground-based
observations. Soon after that, evidence of MSPs have been revealed
in lower mass GGCs, similar to NGC 6752.

A recent paper by Milone et al. (\cite{milonetal10}) is the first
publication devoted to MSPs in NGC 6752. They argue that the
broadening of the MS, visible in the color-magnitude diagram (CMD)
obtained from the high-precision Hubble space telescope photometry
of the cluster cannot be attributed to photometric errors or to
binary stars. Also, a spread in the $(U-B)$ color of RGB stars is
found to correlate with variations in Na and O abundances.

We find and report on direct evidence for the inhomogeneity of the
clusters' stellar population caused by the obvious manifestation,
found for the first time in NGC 6752, of strong radial segregation
between photometrically differing sub-populations of evolutionary
homogeneous cluster stars. We also compare the obtained results with
our recent similar findings in NGC 3201 (Kravtsov et al.
\cite{kravtsovetal10a}) and NGC 1261 (Kravtsov et al.
\cite{kravtsovetal10b}).

\section{The observations and photometric data}
\label{photdat}

In the present study of NGC 6752, we are relying on our new
multi-color $UBVI$ photometry (to be published in a forthcoming
paper) of 18\,508 stars (mainly in the $V$ and $B$) in a
14$\arcmin$x14$\arcmin$ field approximately centered on the cluster,
which reaches a few magnitudes below the turnoff point in all
passbands. The observations were gathered on three consecutive
nights in October 1999, with the 1.3 m Warsaw telescope at Las
Campanas Observatory. A $2048 \times 2048$ CCD camera was used, with
an angular scale of $0\farcs417 {\rm pixel}^{-1}$. We took a total
of 44 frames. The FWHM was estimated to be less than $1\farcs5$ in
21 frames and above $2\farcs0$ in 8 frames. The air mass varied
between 1.204 and 1.514. Despite this modest resolution, MS stars
were measured at a radial distance of up to $\sim1\farcm0$ from the
cluster center, thanks to the proximity of NGC 6752.

NGC 3201 and NGC 1261 were observed with the same facility.
Concerning NGC 3201, in the analysis and plot presented here we used
photometric data corrected for differential reddening and
decontaminated of field stars as described in Kravtsov et al.
(\cite{kravtsovetal09}) and in Kravtsov et al.
(\cite{kravtsovetal10a}). All details concerning our photometry in
NGC 1261 can be found in Kravtsov et al. (\cite{kravtsovetal10b}).

\section{The radial variation of the $U$ level of the SGB}
\label{sgb}

Like in NGC 1261 and NGC 3201, we find the SGB is systematically
brighter in the outer part of NGC 6752.

\begin{figure}
  \centering
 \includegraphics[angle=-90,width=7cm, clip=]{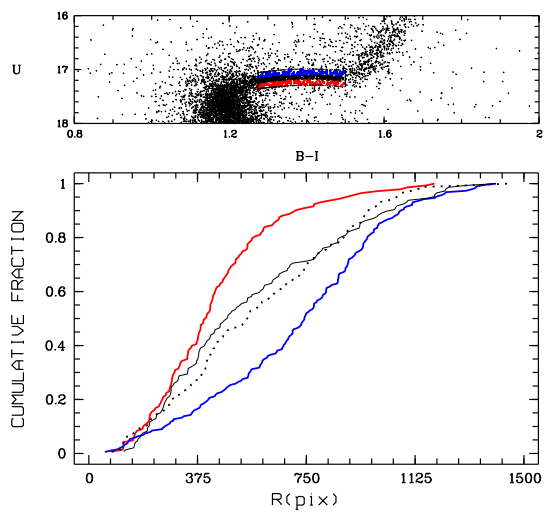}
      \caption{Upper panel: the sub-samples of brighter (blue dots)
and fainter (red dots) SGB stars isolated in the $U$-$(B-I)$ CMD.
Their CRDs are shown by the blue and red lines, respectively, in the
lower panel; these are compared with the CRDs for the sub-samples of
brighter (dotted line) and fainter SGB stars in the $B$ magnitude.
100 pixels correspond to $0\farcm70$.}
         \label{sgbvar}
   \end{figure}

To study the radial variation of the mean level of the SGB in the
$U$-magnitude, we first isolated a total sample of stars most
probably belonging to the SGB. To this aim, we used the $U$-based
CMD with $(B-I)$ color-index, which provides a large separation
between the turnoff (TO) point and the lower RGB. Stars were
selected in the color range $\Delta (B-I) =$ 0.225 ($1.275 < (B-I) <
1.500$) and magnitude range $\Delta U =$ 0.3 mag between the upper
and lower borders of the branch, traced in the CMD and approximated
by two envelope curves. These curves were drawn by fitting the
corresponding SGB envelopes, with second degree polynomials
differing only in their constant term: $U$ = 3.854$(B-I)^2$ -
10.852$(B-I)$ + 24.631 and $U$ = 3.854$(B-I)^2$ - 10.852$(B-I)$ +
24.931 for the upper and lower borders, respectively. The total
number of the selected stars is 451. We then divided the obtained
sample of the SGB stars by three sub-samples: (1) 189 brighter SGB
stars, in a magnitude range $\Delta U =$ 0.12 mag; (2) 143 fainter
SGB stars confined in the same magnitude range; (3) an intermediate
sub-sample of 119 SGB stars falling in the magnitude range $\Delta U
=$ 0.06 in between the two extreme ranges. The selection boxes are
marked in the upper panel of Fig.~\ref{sgbvar}.

The blue and red lines in the lower panel of Fig.~\ref{sgbvar} are
the cumulative radial distributions (CRDs) of the isolated
sub-samples of brighter and fainter SGB stars, respectively, shown
by the same colors in the upper panel. The presented CRDs show that
the fainter sub-giants in the $U$ band are much more centrally
concentrated than the brighter counterparts. These CRDs are
essentially dissimilar, as is supported by a Kolmogorov-Smirnov (KS)
test: they are different at a confidence level of more than 99.9\%.
This notable effect is valid for the $U$ band only. Indeed, in the
lower panel of Fig.~\ref{sgbvar} we plotted for comparison the CRDs
of the sub-samples of brighter (dotted line) and fainter (continuous
line) SGB stars (160 and 143 ones, respectively) in the $B$
magnitude, which were isolated in the same color range as in the
$U$-$(B-I)$ CMD. Brighter and fainter selection boxes are of equal
magnitude ranges, $\Delta B =$ 0.10 mag, separated by a magnitude
range of $\Delta B =$ 0.05 mag. The upper border of the SGB was
accepted at $B = 18.23 - 0.79(B-I)$. The branch is thiner in the $B$
than in the $U$ band. The difference between these distributions is
statistically insignificant, based on a KS test, because they are
different at a confidence level of 75.5\%. This agrees well with a
similar comparison made for SGB stars of NGC 3201 in the $U$ and $V$
magnitudes (Kravtsov et al. \cite{kravtsovetal10a}). Therefore, the
effect we deal with and found for the first time in the three GGCs
(NGC 1261, NGC 3201, and NGC 6752) is detectable only in the $U$
band, but it is absent or (much) less significant in other
photometric bands ($B$, $V$, $I$).

The compared CRDs in the lower panel of Fig.~\ref{sgbvar} imply that
the strong radial segregation of SGB stars in the $U$ magnitude can
hardly be owing to spurious systematic effects caused by a crowding
effect, because this would result in the same spurious effect in all
magnitudes, with opposite dependence. Indeed, systematically
artificially "brighter" SGB stars (blended with fainter MS stars)
are expected to concentrate toward the centers of the clusters, but
not in their outskirts.

\section{The RGB and its color-radial distance diagram}
\label{rgb}

The essential inhomogeneity of SGB stars in NGC 6752 perfectly
agrees with that of RGB stars in terms of a similar difference in
the radial distributions in the cluster of photometrically distinct
sub-populations of red giants.

For our analysis of the RGB of NGC 6752, we selected a sample of the
most probable RGB stars using the advantage of a multi-color
photometry and proceeded like this. Both in the $V$-$(V-I)$ and
$V$-$(B-V)$ CMDs we fitted the mean locus of the RGB with
polynomials applying corresponding commands in the MIDAS system. We
then linearized the RGB by subtracting for each star the color of
the mean locus at its luminosity level from the star's color-index.
We left only those stars that satisfied our selection criterion:
their deviations $\delta(B-V)$ and $\delta(V-I)$ from the mean locus
in both colors simultaneously did not exceed $\pm$ 0.05 mag over the
brighter luminosity range of the RGB ($V < 13.5$) and $\pm$ 0.06
over the fainter range. These conditional boundaries of the RGB in
the color-index separate the bulk of its stars from the majority of
stars belonging to the asymptotic giant branch (AGB) and are
comparable with the mean errors in the colors along the RGB. In this
way we not only rejected the most probable AGB stars and stars
showing a considerable deviation from the RGB fiducial line because
of photometric error, but also minimized the contamination of the
RGB by possible field stars that can appear among the RGB ones in
CMD with a given color-index, but are displaced from the sequence on
CMDs with other color-indices. This selection procedure yielded a
sample of 931 stars.

\begin{figure}
  \centering
 \includegraphics[angle=-90,width=6cm, clip=]{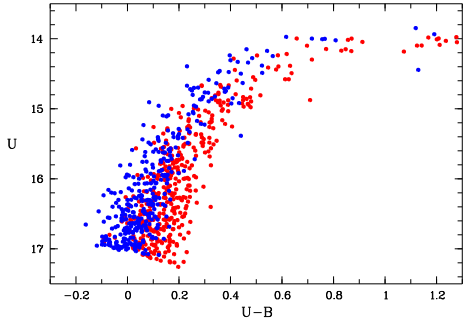}
   \caption{Comparison of the location in the $U$-$(U-B)$ CMD
of RGB stars from the inner (red dots, $0\farcm70 < R <
2\farcm80$) and outer (blue dots, $R > 3\farcm5$) regions of NGC
6752.}
         \label{rgbposit}
   \end{figure}

Based on this sample we first isolated RGB stars located at
different mean radial distances from the center of NGC 6752, namely
in its central ($0\farcm70 < R < 2\farcm80$) and outer ($R
> 3\farcm5$) parts. We plotted these stars in the
$U$-$(U-B)$ CMD with red and blue filled circles, respectively. The
diagram displayed in Fig.~\ref{rgbposit} clearly and unambiguously
shows systematically different location in the $(U-B)$ color of RGB
stars situated at different radial distances from the cluster
center. Evidently RGB stars from the inner region (red circles) are
systematically displaced to the redder $(U-B)$ color compared to
their counterparts from the cluster outskirts. The difference in the
color $(U-B)$ between the two groups of RGB stars was noted by
Marino et al. (\cite{marinoetal08}) in GGC M4. From photometry and
spectroscopy of a sample of 105 stars in the cluster they showed
that the CN-weak red giants with a lower content of Na are on
average systematically bluer, by $\Delta (U-B) = 0.17$ in the
$U$-$(U-B)$ CMD, than CN-strong giants with a higher content of Na.
Regarding NGC 6752, the mean difference in the $(U-B)$ between
reddest and bluest sub-populations of RGB stars can be estimated
using so-called color-position diagram (CPD) discussed below. It is
on the order of $\Delta (U-B) \sim 0.20$, which is approximately a
factor of 1.5 higher than the difference estimated in NGC 3201.

We find CPD is a useful tool to examine in more detail the
dependence, for example, of the $(U-B)$ color of RGB stars on their
radial distance in a cluster. This diagram was originally applied in
our study of the stellar population in the LMC populous star cluster
NGC 1978 (Alca\'ino et al. \cite{alcainoetal99}) and recently for
analysis of inhomogeneous stellar population in NGC 3201 (Kravtsov
et al. \cite{kravtsovetal10a}).

\begin{figure}
  \centering
 \includegraphics[angle=-90,width=6.5cm, clip=]{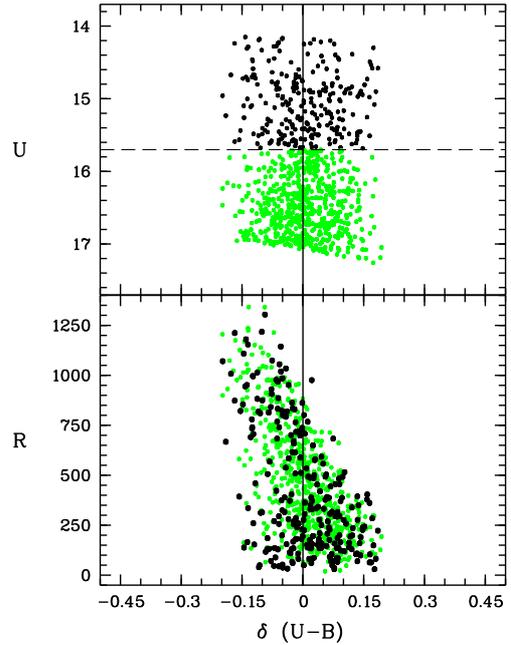}
   \caption{Trend of the $(U-B)$ color of RGB stars with their radial
distance from the center of NGC 6752. Upper panel: the linearized
RGB; its brightest stars ($V < 12.5$) were rejected. For
demonstration purposes, the RGB was arbitrarily divided into two
magnitude intervals at the level ($U = 15.7$) marked by dashed
line, where stars are denoted by filled circles of different
color. The lower panel shows the color-radial distance diagram of
the same RGB stars; 100 pixels correspond to 0\farcm70.}
         \label{cpd}
   \end{figure}

To obtain the CPD of the RGB, we first additionally edited and
slightly cleaned the selected sample of the most probable RGB stars.
Because the brightest part of the RGB, with $V < 12.5$, is
quasi-horizontal in  $U$-$(U-B)$ CMD, we rejected these stars. Then
we proceeded in the same way as described above to linearize the RGB
in the $U$-$(U-B)$ plane and to reject less than two dozen stars
with the largest deviations, $|\delta(U-B)|> \pm 0.20$, from the
mean locus of the RGB. This number is so small compared with the
sample size that this is more a "cosmetic" cleaning with merely
esthetic impact. The linearized RGB is shown in the
$U$-$\delta(U-B)$ plane in the upper panel of Fig.~\ref{cpd}. With
the purpose explained below, the linearized RGB was arbitrarily
divided by two magnitude ranges (at the level marked by the dashed
line) with the following samples of stars: 255 brighter stars with
$U < 15.7$, and 612 fainter stars with $U > 15.7$.

The obtained CPD is shown in the lower panel of Fig.~\ref{cpd}. It
confirms that the $(U-B)$ color does get bluer, i.e. the
$\delta(U-B)$ becomes more negative with increasing radial distance
from the cluster center and that sub-samples of brighter and fainter
RGB stars follow the same trend. However, the CPD is not only
another method to represent the demonstrated radial photometric
inhomogeneity of RGB stars in NGC 6752, but also a plot showing
useful details of this inhomogeneity. Indeed, the CPD assumes the
trend apparently reveals itself beginning at $R \approx 400$ pixels
($2\farcm80$) and there is no obvious trend within this radial
distance where the bulk of stars have positive deviations
$\delta(U-B)$. Interestingly enough, this radial distance is
comparable with the cluster half-mass radius, $R_h = 2\farcm34$
(Harris \cite{harris96}). We noted a very similar behavior of the
CPD of NGC 3201 (Kravtsov et al. \cite{kravtsovetal10a}). We are
inclined to ascribe it to a real effect rather than to some spurious
systematic one for two reasons. If it were some systematic
photometric effect, for example caused by crowding, brighter stars
are normally less affected by it than fainter ones. The behavior of
the CPD traced by the brighter RGB stars (black filled circles in
Fig.~\ref{cpd}) should therefore be more reliable. The second reason
deals with the peculiarity of the radial distribution of both RGB
and SGB stars, which is discussed below.

\begin{figure}
  \centering
 \includegraphics[angle=-90,width=8.0cm, clip=]{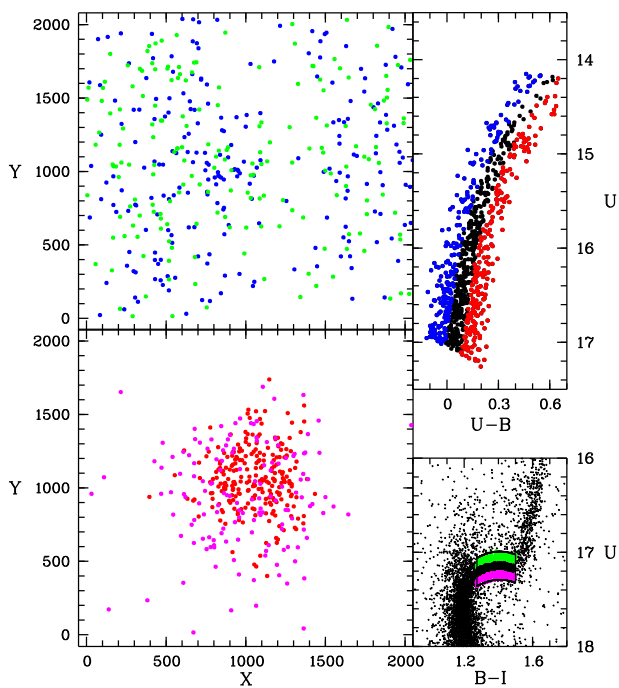}
\caption{Comparison of the location in the observed field of NGC
6752 of (1) two sub-populations of RGB stars, with $\delta(U-B) >
0.05$ (red filled circles) and $\delta(U-B) < -0.05$ (blue filled
circles), and (2) two sub-populations of SGB stars, fainter (magenta
filled circles) and brighter (green filled circles) in the $U$
magnitude. The same stars are shown with the same colors in the
$U$-$(U-B)$ and $U$-$(B-I)$ CMDs of the RGB and SGB, respectively.
The X and Y coordinates increase to the east and north.}
         \label{rgbsgbxy6752}
   \end{figure}

\begin{figure}
  \centering
 \includegraphics[angle=-90,width=8.0cm, clip=]{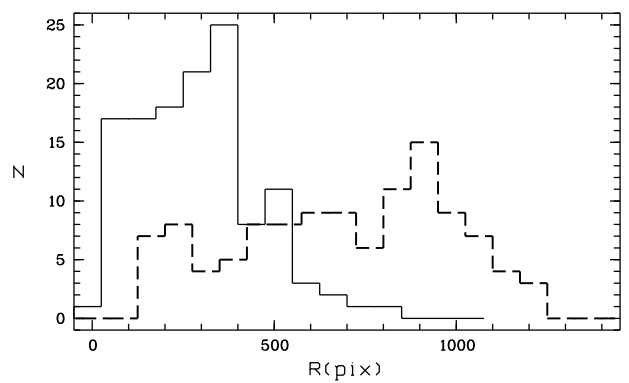}
\caption{Radial distributions of RGB stars with a different
deviation $\delta(U-B)$ from the ridgeline of the RGB in the
$U$-$(U-B)$ diagram of NGC 3201. The continuous and long-dashed
lines are the histograms of RGB stars with $\delta(U-B)
> 0.05$ and $\delta(U-B) < -0.05$, respectively. 100 pixels
approximately correspond to $0\farcm70$.}
         \label{rgbradist3201}
   \end{figure}

We compared the distribution of photometrically distinct
sub-populations of SGB and RGB stars in the observed field of NGC
6752. With this purpose we selected the sub-populations of "blue"
and "red" RGB stars with $\delta(U-B) < -0.05$ and $\delta(U-B) >
0.05$, respectively. This separation (conditional to a certain
extent, of course) minimizes the mutual contamination between the
two sub-populations by stars with large photometric errors. Two
sub-populations of the SGB are the described above samples of
brighter and fainter stars in the $U$ magnitude. We show the
resulting four sub-samples of stars by different colors in
respective CMDs and in the cluster field in Fig.~\ref{rgbsgbxy6752}.
Blue and red filled circles are the selected "blue" and "red" RGB
stars, while green and magenta filled circles are brighter and
fainter SGB stars in the $U$ band, respectively. This figure clearly
shows a strong difference in the radial distribution between the
sub-populations of the "blue" and "red" RGB stars. Moreover, the
same very similar difference between the radial distribution of SGB
stars brighter and fainter in the $U$-magnitude is obvious in NGC
6752 as well. On the other hand, the faint SGB and red RGB stars are
similarly much more centrally concentrated than their respective
bright and blue counterparts. Interestingly enough NGC 3201 exhibits
virtually the same very similar behavior of the radial distribution
of photometrically distinct sub-populations of SGB and RGB stars.
The distribution of the latter stars is shown in
Fig.~\ref{rgbradist3201}.

Fig.~\ref{rgbsgbxy6752} shows that the distribution of stars in the
X,Y planes, especially of the "bluer" RGB and "brighter" SGB stars
distributed across a larger area, is somewhat spotted, and the
centers of the distributions of sub-populations in the upper and
lower plots are somewhat different. These effects are probably
caused by a combination of different factors. To name a few major
factors: (i) statistical fluctuation caused by the relatively small
"surface density" of the "bluer" RGB and "brighter" SGB stars and
(ii) larger confusion (poorer distinguishing) between the
sub-populations of the same evolutionary sequence toward the center
of the cluster, caused by increased systematic and random
photometric errors.

\section{Conclusions}

Based on a new multi-color photometry in a fairly wide field of
NGC 6752, we found and report the following direct evidence of the
inhomogeneity (multiplicity) of the cluster's stellar population.

There is an essential radial segregation of SGB stars in the
cluster, depending on their brightness in the $U$ band: the fainter
sub-population of sub-giants is obviously centrally concentrated,
and the bulk of these are confined within the cluster region with
the half-mass radius, $R_h$, while their brighter counterparts show
a flatter radial distribution within the observed field.

The sub-populations of photometrically distinct RGB stars also
exhibit essential radial segregation in NGC 6752: RGB stars redder
in the color $(U-B)$, like the sub-population of the fainter SGB
stars, are centrally concentrated mostly within the region with
radius $R \sim R_h$, as opposed to the sub-populations of both RGB
stars bluer in the $(U-B)$ color and SGB stars brighter in the $U$
band. We note virtually the same radial segregation between
photometrically distinct sub-populations both of RGB and SGB stars
in NGC 3201. In NGC 1261 (Kravtsov et al. \cite{kravtsovetal10b}),
we also found similar trends, but we are not able to adequately
judge what exactly happens in its central part.

The revealed and demonstrated strong radial segregation between the
sub-populations in NGC 6752 and NGC 3201 is closely related with the
difference of their photometric characteristic either in the $U$
magnitude (SGB stars) or in the $(U-B)$ color (RGB stars), which are
known to be sensitive to metallicity. Therefore, the obtained
results not only provide evidence for the radial segregation itself
between the sub-populations, but also imply (1) the radial
dependence of the abundance of those elements, which are mainly
responsible for this photometric difference between stars in the
clusters, and (2) presumably different kinematics between the
sub-populations. In this context, it would be important to study
whether there is a radial trend (and what it is exactly) of the
elemental abundance in NGC 6752 and other GGCs.

\begin{acknowledgements}

We thank the anonymous referee for useful comments that improved the
manuscript.

\end{acknowledgements}

\end{document}